\documentstyle[prabib,aps,twocolumn,prl,epsfig]{revtex}

\begin{document}
\draft
\preprint{HEP/123-qed}
\title{Morphology and thermal conductivity of model organic aerogels}
\author{Anthony P. Roberts\cite{byline}}
\address{
Faculty of Environmental Sciences,
Griffith University,
Nathan,
Queensland, 4111 Australia}
\date{\today}
\maketitle
\begin{abstract}
The intersection volume of two independent 2-level cut Gaussian random
fields is proposed to model the open-cell microstructure of organic
aerogels.
The experimentally measured X-ray scattering intensity, surface area
and solid thermal conductivity of both {\it polymeric} and
{\it colloidal} organic aerogels can be accounted for by the model. 
\end{abstract}

\narrowtext
Aerogels are a promising material for a host of
applications~\cite{Fricke86,Fricke92} due to their
thermal, optical and mechanical properties. 
For example, aerogels are among the best thermal insulating
solid materials known~\cite{Lu92,Hrubesh94,Fricke91}.
It is important to link aerogel properties to their
complex internal microstructure, and to understand
how such properties can be optimized for a given
application~\cite{Fricke92,Hrubesh94,Lu95}.
The nano-scale porous morphology of aerogels has been
extensively characterized by X-ray scattering and surface area
analysis~\cite{Schaefer84,Pekala93,Schaefer94,Hasmy94}.
Despite this, aerogel properties are usually
correlated with density, rather than related to morphological
features. One reason for this is
the lack of suitable representation of aerogel morphology.
In this paper we develop a model which accounts for
the open-cell morphology of organic aerogels.
The solid thermal conductivity of the model is computed and
shown to be in good agreement with experimental data.

Thermal transport in aerogels is due to three additive components:
conduction in the solid skeleton and (gas-filled) pores and
conduction due to radiation~\cite{Lu92}.
For thermal insulation purposes it is desirable to reduce the
magnitude of each contribution.  Gaseous conductivity can be
significantly reduced by decreasing pore size or partially evacuating
the material, and radiative transport reduced by the inclusion of
an opacifier~\cite{Lu92,Hrubesh94}.
The solid conductivity (typically half the total)
depends strongly on the aerogel density and
microstructure~\cite{Hrubesh94,Lu95}.

Organic aerogels produced by the polymerisation of resorcinal
and formaldehyde (RF) have been suggested as an alternative
insulator to opacified silica aerogels~\cite{Lu92,Pekala89}.
They have lower intrinsic and
radiative conductivities, and are less brittle than their silica based
counterparts~\cite{Lu92,Hrubesh94}.
Both the morphology~\cite{Pekala93,Schaefer95a} and properties
of organic aerogels~\cite{Lu92,Lu95,Pekala90,Reynolds94,Scherer96a} have
been the subject of detailed investigation.        
A key variable in the formation of RF aerogel microstructure is
the initial ratio of resorcinal to
catalyst (R/C)~\cite{Pekala93}. 
As the catalyst increases the aerogels
vary from a colloidal structure to a well-connected polymeric
structure with a corresponding increase in conductivity and
strength~\cite{Lu95,Pekala90}.
It is important to quantitatively model these properties to
assist in the understanding and optimization of RF aerogels.

Current models of aerogels are based on simulating
the microstructure formation using the diffusion-limited
cluster-cluster aggregation (DLCA)
scheme~\cite{Hasmy94,Meakin83c,Kolb83,Chandler95}. Two features
of DLCA models (proposed for silica aerogels) suggest 
that they are not well suited to modeling RF aerogels.   
Firstly, the DLCA model exhibits
fractal scaling~\cite{Meakin83c,Kolb83} and a well pronounced peak
in the scattering intensity~\cite{Hasmy94,Chandler95}.
In contrast, RF aerogels exhibit no fractal scaling, and under high catalyst
conditions the peak is weak, or even absent~\cite{Pekala93}.
Secondly, the discrete character of DCLA type models (open networks
of cubes or hard spheres) may be ill-suited to modeling
``continuum" properties within the aerogel skeleton.
For example, the influential inter-particle neck size~\cite{Reynolds94}
is equal to zero for hard spheres~\cite{Hasmy94}, and equal to
the particle size for cubes~\cite{Meakin83c,Kolb83,Chandler95}.
We propose a statistical model of microstructure
which can account for the main morphological features of RF aerogels.
The model is lattice independent, and suitable for
continuum based theoretical and computational
prediction of properties~\cite{TorqRev91,Roberts95a}.

A convenient statistical description of porous media
is provided by modeling the internal interface as the iso-surface
(or level-cut) of a Gaussian random field (GRF) $y(\mbox{\boldmath$r$})$.
This model has been used to describe the morphologies arising in
spinodal decomposition~\cite{Cahn65},
microemulsions~\cite{Marcelja90,Teubner91}, galaxy
formation~\cite{Politzer84} and porous rocks~\cite{Quiblier84}
amongst others~\cite{Isichenko92}.
The statistics
of the material are completely determined by the specification of
the single level-cut parameter and the field-field correlation function
$g(r)=\langle y(\mbox{\boldmath$0$})y(\mbox{\boldmath$r$}) \rangle$
(where $r=|\mbox{\boldmath$r$}|$ and $g(0)\equiv 1$).
Berk generalized the model to account for the X-ray scattering
properties of microemulsions~\cite{Berk87} by defining
phase 1 to occupy the region in space where
$\alpha\leq y(\mbox{\boldmath$r$}) \leq \beta$ and phase 2 to occupy the
remainder.  This model has also been shown to account for
the morphology and properties of foamed solids~\cite{Roberts95b}
and polymer blends~\cite{Knackstedt95a}. 

Neither the 1-cut GRF model, or Berk's ``2-cut'' extension, can account for the high 
porosity open-cell microstructure of aerogels. The 1-cut GRF is not
macroscopically connected at aerogel porosities~\cite{Roberts95a}
(typically 95\%), and Berk's 2-cut model exhibits sheet-like
structures~\cite{Roberts96a} similar to those observed in
closed-cell foams~\cite{Gibson88}.  To model the open-cell microstructure
we define the solid phase to occupy the region
$\alpha \leq y(\mbox{\boldmath$r$}) \leq \beta$ {\it and}
$\alpha\leq w(\mbox{\boldmath$r$}) \leq \beta$
where $y$ and $w$ are statistically independent GRF's.
The independence of the random fields allows the correlation
functions of the model to be calculated.
The solid volume fraction of the model is just
$p=(p_\beta-p_\alpha)^2$  where
\mbox{$p_\alpha=(2\pi)^{-\frac12}\int_{-\infty}^\alpha e^{-t^2/2} dt$}
and the 2-point correlation function is $p_2(r)=h^2_2(r)$ where
$h_2(r)$ is the 2-point function of Berk derived for the same values
of $\alpha$ and $\beta$ 
\begin{eqnarray}
h_2(r)= && h^2+\frac{1}{2\pi}\int_0^{g(r)}
 \frac{dt}{\sqrt{1-t^2}} \times  \left[
\exp\left({-\frac{\alpha^2}{1+t}}\right) \right.  
\label{h2}
\\ && \left. \nonumber
-2\exp\left({-\frac{\alpha^2-2\alpha\beta
t+\beta^2}{2(1-t^2)}}\right)
+\exp\left({-\frac{\beta^2}{1+t}}\right) \right]
\end{eqnarray}    
with $h=p_\beta-p_\alpha$.
The freedom in specifying the level-cut parameters and the field-field
correlation function of the model $g(r)$ allow a wide variety
of morphologies to be modelled.

To relate the model to experimental data it is necessary to
specify a field-field correlation function. Prior
studies~\cite{Marcelja90,Teubner87} suggest a form
\begin{equation}
g(r)=\frac{e^{-r/\xi}-(r_c/\xi)e^{-r/r_c}}{1-(r_c/\xi)}
\frac{\sin 2\pi r /d}{2\pi r /d}
\label{gr}
\end{equation}
characterized by a correlation length $\xi$, domain scale $d$ and a cut-off
scale $r_c$.
Two commonly measured morphological quantities of porous media are
the surface area $S$ and the X-ray scattering
intensity $I(q)$. These can be computed
for the model as
\begin{equation}
\frac SV = \frac{4\sqrt{2p}}{\pi} \left( e^{-\frac12 {\alpha^2}}+
e^{-\frac12 \beta^2} \right)\sqrt{\frac{4\pi^2}{6d^2}+\frac{1}{2r_c\xi}}
\end{equation}
and
\begin{equation}
\frac{I(q)}V=\langle\eta^2\rangle\int_0^\infty4\pi r^2 (p_2(r)-p^2)
\frac{\sin qr}{qr} dr,
\label{scatter}
\end{equation}
where $V$ is the sample volume and $\eta$ is the scattering density 
of the solid phase~\cite{Debye57}.

\begin{figure}[bt!]
\centering\epsfig{figure=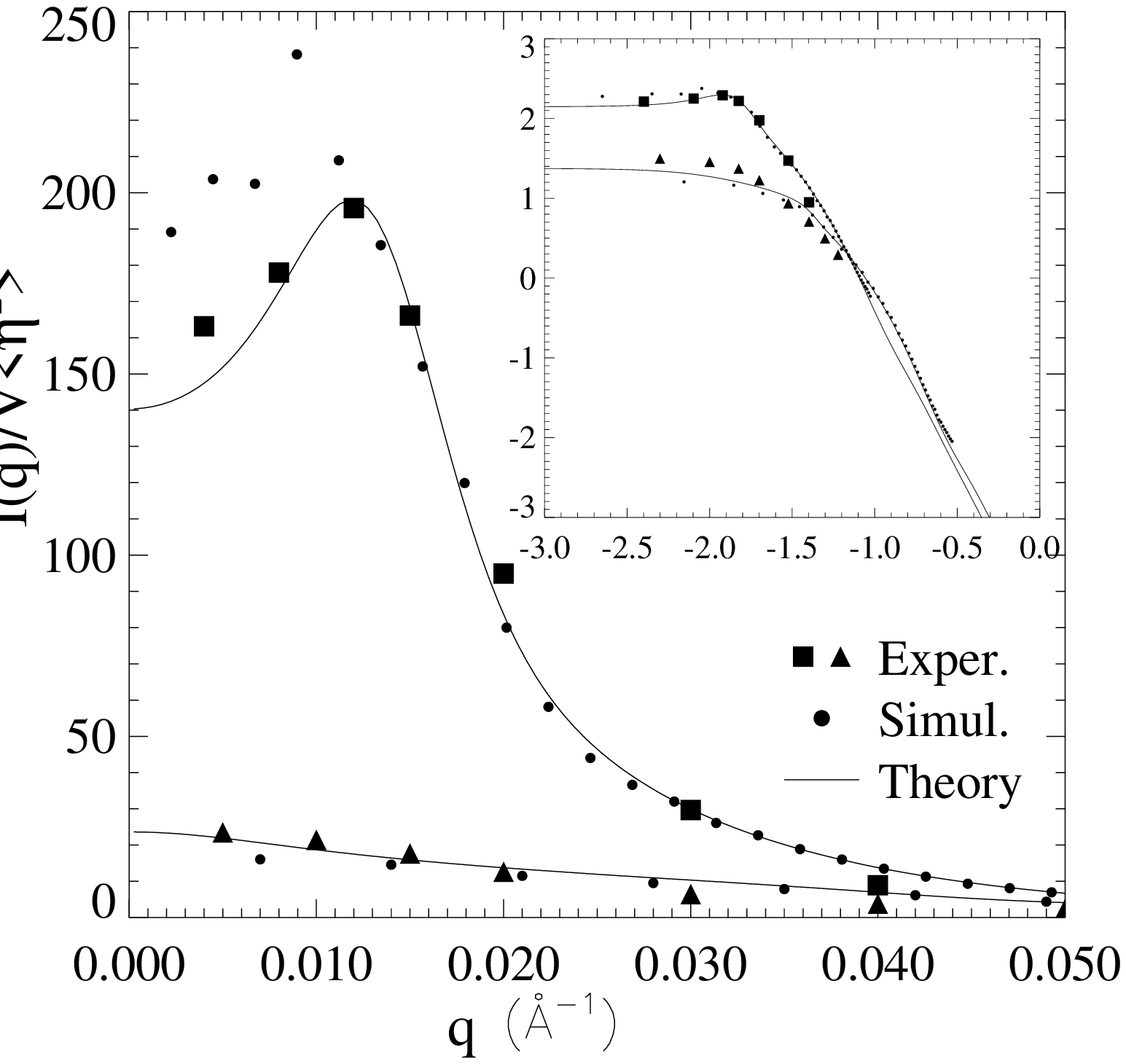,width=8.0cm} 
\caption{X-ray scattering spectra of RF aerogels.
The data for the upper and lower curves are for a
colloidal (R/C=300, $p$=11\%) and  polymeric (R/C=50, $p$=7.7\%) aerogel
respectively~\protect\cite{Pekala93} (the experimental data
are vertically scaled).  The theoretical curves were obtained
using Eq.~(\ref{scatter}), and the simulation data are measured
directly from one realization
of each model (Figs.~\ref{114p11}~\&~\ref{123p08}). The inset
shows $\log_{10}(I(q)/V\langle\eta^2\rangle)$ vs.\ $\log_{10}(q)$.
\label{xray2}} 
\end{figure}

\begin{figure}[bt!]
\centering\epsfig{figure=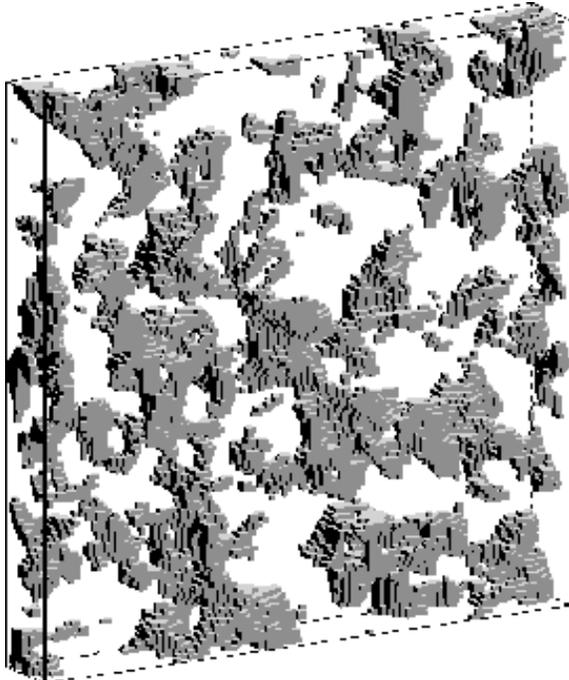,width=7.5cm} 
\caption{A model {\it colloidal} aerogel (solid fraction $p=$11\%).
The parameters are $r_c$=10, $\xi$=14, $d$=46nm and
$p_{\alpha,\beta}=0.07,0.40$. The image is $276\times276\times34.5$nm.
The slab is part of a periodic cubic sample of side
length 276nm (128 pixels).  Many of the apparently isolated
clusters are interconnected outside the volume shown.
\label{114p11}}
\end{figure}

To model RF aerogels we choose the model parameters to match
experimentally measured scattering and surface area
data~\cite{Pekala93}. While the domain scale
$d$ corresponds to the pore scale in aerogels, the geometry of
the fibres depends on both the length scale and level-cut parameters.
Uncertainties in the estimation of surface
area~\cite{Schaefer94,Schaefer95a,Emmerling92} and skeletal
density of aerogels suggest that only rough approximations of the
parameters are justified.
Examples of a colloidal and polymeric aerogel are chosen to
ascertain the generality of the model.
The colloidal aerogel is produced under low catalyst
conditions (R/C=300) and has
density $\rho_a=148$kgm$^{-3}$ ($p=\rho_a/\rho_s=0.11$)~\cite{Pekala93}.
TEM images show a ``string-of-pearls" appearance:
a network comprised of grains (diameter 120-300\AA)
interconnected by narrow necks. The surface area is $400$m$^2$/g
and X-ray scattering yields a peak at $q=0.012$\AA$^{-1}$
(Fig.~\ref{xray2})~\cite{Pekala93}.
A good match between the experimental information and the model
was obtained for $r_c=10,\xi=14$ \& $d=46$nm and
$p_{\alpha,\beta}=.07,.40$.
The model surface area is $428$m$^2$/g, and the theoretical scattering
curve is seen to be in good agreement with the experimental data
(Fig.~\ref{xray2}). In Fig.~\ref{114p11} we show a slab of the 
material: the model reproduces the colloidal string-of-pearls
morphology with reasonable grain and neck sizes.
We directly measure the scattering from the model.
The results are included in Fig.~\ref{xray2}, demonstrating
that the simulation reproduces the theoretical
statistics very well.  At small $q$ (large length scales)
some deviation is evident; this is due to the finite size of
the samples~\cite{RobertsUP}.

\begin{figure}[bt!]
\centering\epsfig{figure=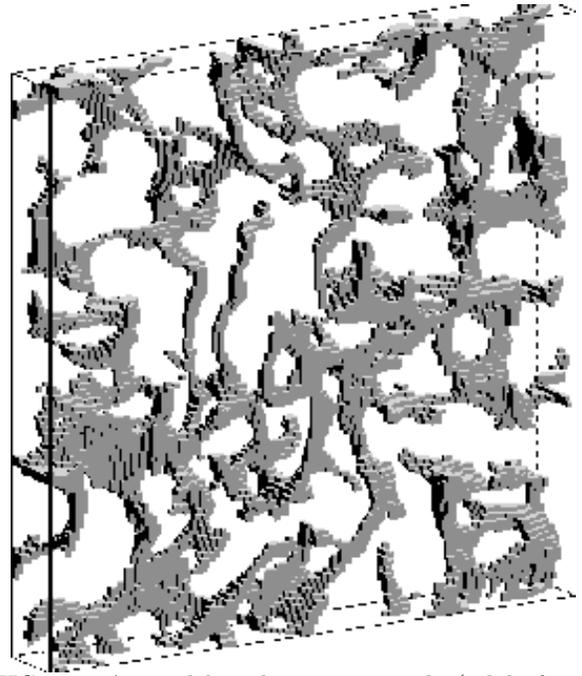,width=7.5cm} 
\caption{A model {\it polymeric} aerogel (solid fraction $p=7.7$\%).
The model parameters are $r_c$=10, $\xi$=20, $d$=30nm and
$p_{\alpha,\beta}=0.36,0.64$. The image is $90\times90\times11.25$nm.
\label{123p08}} 
\end{figure}

The polymeric aerogel~\cite{Pekala93} we model is produced under
high catalyst conditions (R/C=50) and has
density $\rho_a=100$kg/m$^{3}$ ($p=0.077$). 
The aerogel exhibits a network of uniform fibres
(diameter 30-60\AA) with surface area $905$m$^2$/g.
The scattering intensity monotonically decreases with $q$ and yields
a Guinier radius of $81$\AA~\cite{Pekala93}.  
In this case we take $r_c=10,\xi=20$ \& $d=30$nm
and centered level cut parameters ($\alpha=-\beta$)
$p_{\alpha,\beta}=.361,.639$.  The model has a surface area of
$927$m$^2$/g and a Guinier radius of 101\AA.
Fig.~\ref{xray2} shows good agreement between the model and
experimental scattering curves. 
A slab of the model material is shown in Fig.~\ref{123p08}. The fibres 
have a relatively uniform thickness, varying from $20-60$\AA.

As the presence of a peak in the scattering may yield information
about the the physical processes underlying aerogel
formation~\cite{Pekala93,Schaefer95a} it is interesting to
comment on its morphological origins.
The existence of a domain (or repeat) scale in a
random structure leads to decaying oscillations in the
correlation function, and hence a peak in $I(q)$.
In aerogels the decay scale is controlled by the width of
the fibres $w_f$, and the domain scale $d$ is that of the pores. 
If $d$ is only several times larger than $w_f$
(e.g.\ the colloidal model) a peak is
observed. On the other hand, if $d$
is an order of magnitude larger than $w_f$
(as it is in the polymeric model) the oscillations
in $p_2(r)$ are smoothed by a stronger decay and the
peak is extinguished (Fig.~\ref{xray2}). Note that pores
with a well defined scale are evident in the
model (Fig.~\ref{123p08}); they simply do not carry sufficient
statistical weight to appear in the scattering.

We now compare the thermal conductivity of the model to experimental data.
The {\it solid} thermal conductivity of RF gels has been
experimentally measured 
over the density range $\rho_a=60-400$kgm$^{-3}$ at catalyst
concentrations (R/C=50,200 \& 300)~\cite{Lu92,Lu95}.
To estimate the model conductivity we assume that the local heat
flux obeys the Fourier law
$\mbox{\boldmath$j$}=-\lambda(\mbox{\boldmath$r$})
\mbox{\boldmath$\nabla$} T$ where
$\lambda(\mbox{\boldmath$r$})=\lambda_s(0)$ in the solid(void) phase.
Conventional numerical techniques are used to solve the heat conservation
equation $\mbox{\boldmath$\nabla$}
\cdot(\lambda\mbox{\boldmath$\nabla$}T)=0$ in a 128$^3$ lattice
subject to an applied temperature gradient~\cite{Roberts95a}.
The aerogel conductivity is obtained as
$\lambda_a=\langle \lambda(\mbox{\boldmath$r$})\mbox{\boldmath$\nabla$}
T\rangle/ \langle \mbox{\boldmath$\nabla$} T \rangle$.

\begin{figure}[bt!]
\centering\epsfig{figure=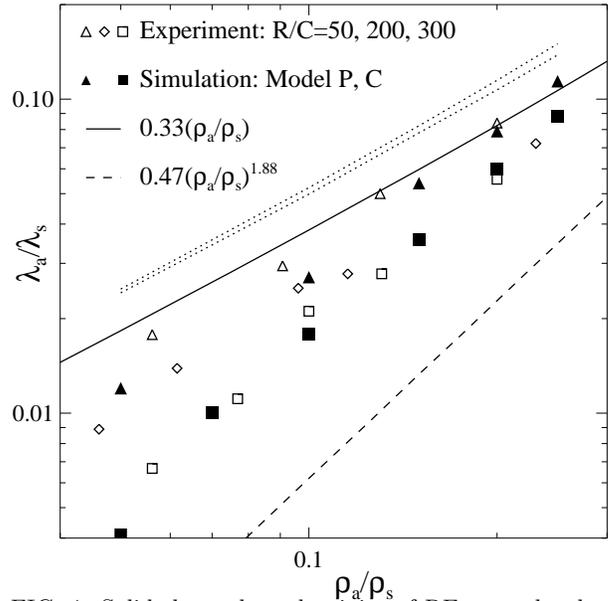,width=8.0cm} 
\caption{Solid thermal conductivity of RF aerogels;
theory vs.\ experiment~\protect\cite{Lu95}
($\rho_s=1300$kg/m$^3$, $\lambda_s=0.18$Wm$^{-1}$K$^{-1}$).
The estimates of $\lambda_a$ from this work (solid symbols) show very
good agreement with experimental data (open symbols). The solid and
dashed lines correspond to predictions of $\lambda_a$ discussed
in the text. The dotted lines are rigorous upper bounds.
\label{hru1}} 
\end{figure}

Note that we have derived models of colloidal and polymeric 
aerogels based on the morphology and scattering data
at a specific density. We can extend the model to higher
and lower densities by making simple assumptions about the
density dependence of the model parameters. A simple scaling argument
shows the pore scale varies as $d\propto p^{-\frac12}$~\cite{Hrubesh94}
and for simplicity a similar dependence is assumed for $r_c$ and $\xi$
(the conductivity is relatively insensitive to length scale
variations~\cite{Roberts95a,Roberts96a}).
The polymeric morphology of the high catalyst (R/C=50) aerogels
was reproduced by ``centering" the level cut
parameters $(\alpha=-\beta)$ (Fig.~\ref{123p08}).
We preserve this feature of the model by choosing
$p_\beta=\frac12+\frac12 \sqrt p$ and $p_\alpha=\frac12-\frac12 \sqrt p$.
The thermal conductivity of the polymeric model ``P" is shown
in Fig.~\ref{hru1}. While slightly underestimating the conductivity of
the aerogel produced at R/C=50, it is nevertheless seen to be
in very good agreement with the experimental data.
At low catalyst concentration (R/C=300) the aerogel morphology
was modeled by non-centered level cuts $(\alpha \neq -\beta)$
(Fig.~\ref{114p11}). To maintain an asymmetry we choose
$p_\beta=0.3+\frac12 \sqrt p$ and $p_\alpha=0.3-\frac12 \sqrt p$.
The thermal conductivity of this model ``C" is presented
in Fig.~\ref{hru1}, providing excellent agreement with the experimental
data for the RF aerogel produced at R/C=300. 

In Fig.~\ref{hru1} we have also plotted a number of results
arising from theoretical considerations. Zeng {\it et al}~\cite{Zeng95}
have suggested that periodic open-cell models can be used
to estimate aerogel conductivity. 
At low relative densities the
``square rod" model leads to the estimate
$\lambda_a/\lambda_s=\frac13 \rho_a/\rho_s$~\cite{Roberts96a}
in remarkably good agreement with the data for polymeric RF aerogels. 
From considerations of phonon heat transport in solids
it has been suggested that $\lambda_a/\lambda_s=\rho_a v_a/\rho_sv_s$
where $v_a(v_s)$ is the sound velocity in the
aerogel(solid)~\cite{Nilsson89}.
Measurements performed on RF aerogels have determined that
$v_a/v_s=0.47 (\rho/\rho_s)^{0.88}$ so that
$\lambda_a/\lambda_s=0.47(\rho_a/\rho_s)^{1.88}$~\cite{Hrubesh94,Gross92}.
The result is seen to
significantly under-estimate the measured conductivity of RF aerogels.
It is also possible to calculate rigorous variational
bounds~\cite{Beran65a,Milton81a} on the model thermal
conductivity~\cite{Roberts95a,Roberts96a,RobertsUP}. 
The upper bounds, which can have predictive
power~\cite{TorqRev91,Roberts96a}, are seen to considerably
over-estimate the true conductivity.
Thus theoretical microstructure-property relations are unable
to predict the thermal conductivity, and numerical simulations
must be relied on.

The agreement between the model and experimental data for both
colloidal and polymeric aerogels provides strong evidence that 
we have accurately modelled the morphology of organic aerogels.
The results also indicate that Fourier's continuum theory of
heat conduction may hold even in nano-scale structures
(diameter 30-60\AA).
Of course there is no guarantee that the model is correct: other
models may share the same morphological~\cite{Marcelja90}
and thermal properties.
Nevertheless the utility of the model has been shown.
It should be possible to apply the model in the study of gas and radiative
conductivity and the mechanical properties of aerogels.
The fractal properties of silica aerogels can
also be incorporated~\cite{RobertsUP}.

Extensions of the model are relevant to a
wider range of heterogeneous materials.
For example, the solid phase of the aerogel model
mimics the inter-granular pores of sandstone,
and micro-porosity can be simulated by including
random structures at smaller scales.
Spheres~\cite{TorqRev91} may be embedded in the models,
and closed cell morphologies, such as those observed
in solid foams~\cite{Gibson88}, can be formed from
the {\it union} set of two 2-level cut GRFs. 
The model correlation functions can be calculated,
allowing surface areas, scattering curves and
rigorous property bounds to be evaluated~\cite{RobertsUP}.


\clearpage

\psfull
\end{document}